\documentclass[showpacs,floatfix,superscriptaddress,showpacs,preprint,amssymb,amsfonts,prb,aps]{revtex4}
\usepackage{longtable,graphicx,epsfig,dcolumn}

\begin{document}
\bibliographystyle{revtex}
\title{Microscopic Surface Structure of Liquid Alkali Metals }
\author{Holger~Tostmann}
\affiliation{Division of Applied Sciences and Department of Physics,
Harvard University, Cambridge MA 02138}

\author{Elaine~DiMasi}
\affiliation{Department of Physics, Brookhaven National Laboratory,
Upton NY 11973-5000}

\author{Oleg~G.~Shpyrko}
\affiliation{Division of Applied Sciences and Department of Physics,
Harvard University, Cambridge MA 02138}

\author{Peter~S.~Pershan}
\affiliation{Division of Applied Sciences and Department of Physics,
Harvard University, Cambridge MA 02138}

\author{Benjamin~M.~Ocko}
\affiliation{Department of Physics, Brookhaven National Laboratory,
Upton NY 11973-5000}

\author{Moshe~Deutsch}
\affiliation{Department of Physics, Bar-Ilan University, Ramat-Gan
52100, Israel}

\begin{abstract}

We report an x-ray scattering study of the microscopic structure of
the surface of a liquid alkali metal. The bulk liquid structure
factor of the eutectic K$_{67}$Na$_{33}$ alloy is characteristic of
an ideal mixture, and so shares the properties of an elemental
liquid alkali metal. Analysis of off-specular diffuse scattering and
specular x-ray reflectivity shows that the surface roughness of the
K-Na alloy follows simple capillary wave behavior with a surface
structure factor indicative of surface induced layering. Comparison
of thelow-angle tail of the K$_{67}$Na$_{33}$ surface structure
factor with the one measured for liquid Ga and In previously
suggests that layering is less pronounced in alkali metals.
Controlled  exposure of the liquid to H$_2$ and O$_2$ gas does not
affect the surface structure, indicating that oxide and hydride are
not stable at the liquid surface under these experimental
conditions.
\end{abstract}


\maketitle
\section{Introduction}

The structure of the free surface of a liquid metal (LM)
is fundamentally
different from that of a dielectric liquid, due to the strong
coupling between conduction electrons and ion cores.\cite{harr87}
At the liquid-vapor interface of a LM,
the Coulomb interaction between the
free electron Fermi gas and the classical gas of charged ions
acts like an effective hard wall and forces the ions into
ordered layers parallel to the surface.
The existence of surface-induced layering in  LM  has  been
verified unambiguously by experiment  for liquid
Hg\cite{mercury}, Ga\cite{gallium} and In.\cite{indium}
Layering is also present in liquid binary alloys,
even though it may be suppressed by surface segregation\cite{gaprb}
or the formation of surface phases.\cite{bunsen,srn}

Even though surface layering appears to occur in LM in general,
comparison of the detailed surface structure
reveals qualitative differences between different
LM.\cite{indium,mercurytemp}
In fact, very little is understood about how the
surface induced layering
predicted for an ideal LM
is affected by the details of the more complicated electronic  structure
exhibited by the polyvalent metals studied so far.
One important goal therefore is to study the surface structure
of those LM which show the most ideal  electronic structure
in the bulk phase, characterized by itinerant conduction electrons
which are only weakly perturbed by a small
ionic pseudopotential. These LM are
referred to as nearly-free electron or NFE-type liquid
    metals.\cite{ziman61}
In a previous study we have compared the surface structure of
liquid In, which is considered to be a NFE-type
liquid metal, with the surface structure of liquid Ga, which displays
a considerable degree of covalency in the
    bulk.\cite{indium}
However, even though liquid In has less tendency towards
covalent bonding than Ga,
it is trivalent, and the most simple models do not adequately
describe its electronic structure or metallic
    properties.\cite{ashcroft}
The most ideal NFE-type liquid metals are the monovalent alkali
metals (AM) which  have a half-filled $s$-band and
an almost spherical Fermi
    surface.\cite{march}
The metallic properties of the AM are explained by the most
simple solid state models such as the Drude
    theory.\cite{ashcroft}
The fact that the conduction band is primarily formed
from  $s$-electron states
and  that the electron-ion interaction can be modeled by a
weak pseudopotential have led to numerous
theoretical studies on the electronic structure.
In fact, as far as the surface structure
of LM is concerned, most of the theoretical studies and
computer simulations have been applied to liquid Na, K and
    Cs.\cite{harr87,alkalitheory}

A particularly interesting question concerns the possible correlation
between surface tension and surface structure. In general, LM  have
surface tensions that are at least an order of magnitude
higher than the surface tensions of all other types of
    liquids.\cite{srn}
Liquid alkali metals are the only exception to this pattern.
For example, Ga and Cs melt at about the same temperature
 but the surface
tension of Ga is greater than that of Cs by about an order of magnitude.
If the high surface tension of
liquid metals is related to surface induced layering, it would
be reasonable to expect that liquid AM should display weak
or no layering at the surface. This is contradicted by
computer simulations showing pronounced layering for liquid
    Cs.\cite{harr87}

\begin{figure}[tbp]
 \epsfig{file=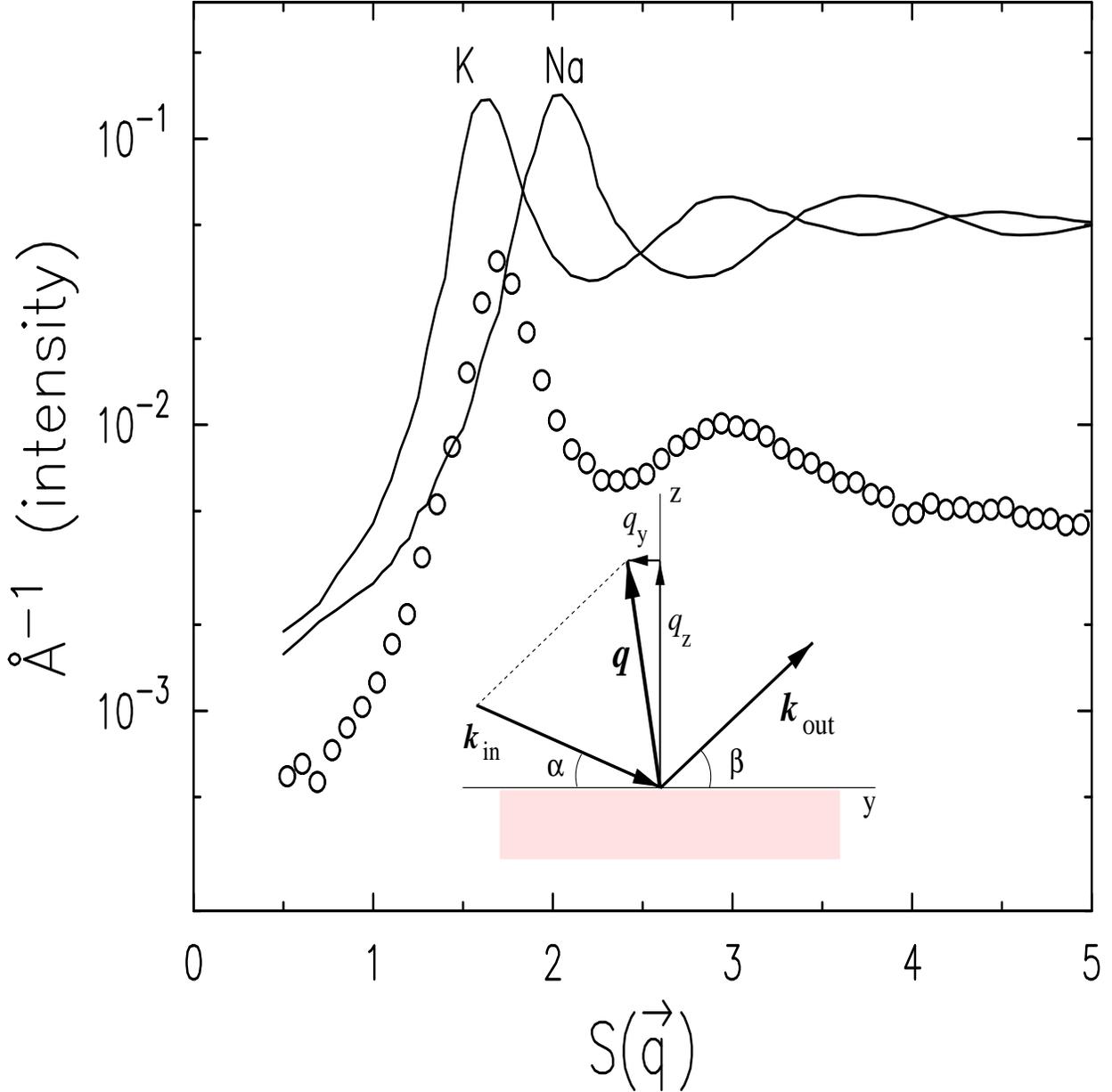, angle=0,
width=1.0\columnwidth} \caption{ Bulk liquid structure factor
measured
 for liquid K$_{67}$Na$_{33}$ ($\circ$)
compared to the  bulk structure factors of pure liquid K and Na from
Ref.~14 (---). Inset: geometry for x-ray reflectivity and
off-specular diffuse scattering as described in the text.
}\label{kna_fig:phase}
\end{figure}

The alkali alloy  K$_{67}$Na$_{33}$ is more suitable
experimentally for such a study than are the elemental
liquid AM.
Among the elemental alkali metals,  Li deviates from NFE-type
    behavior,\cite{ashcroft,addison}
while the high  equilibrium vapor pressures
of the other elemental AM prevents the use of an
 ultra-high vacuum (UHV) environment over a long
period of time without noticeable evaporation of material.
Experiments under  UHV conditions are mandated by the high
reactivity of liquid AM towards water and oxygen.\cite{addison}
When liquid K and Na are alloyed in the eutectic composition
(67~at\%K), however, a vapor pressure well below $10^{-10}$ Torr
is obtained. The low melting point of this alloy ($-12.6^\circ$C)
is an additional advantage, since the thermal surface roughness is
considerably reduced.\cite{indium} This binary alloy appears to be
well  suited for emulation of  the properties of an  elemental
liquid AM. The total structure factor $S(q)$ of bulk liquid
K$_{67}$Na$_{33}$, shown in Figure~\ref{kna_fig:phase}, is
characteristic of random mixing, without subpeak or asymmetry of
the first peak.\cite{waseda} The first peak of $S(q)$ for the
alloy lies between those of pure  K and Na, as expected for an
ideal
    mixture.\cite{waseda}
More importantly, the electronic
structure of the homovalent K-Na alloys is as NFE-like as the
electronic structure of the pure
    components.\cite{shimoji}
Even if there were some homocoordination or surface enrichment of
K driven by  the  lower surface tension of K as compared to Na, it
would not be discernible in our structural measurements, since
x-ray scattering is sensitive to the distribution of the electron
density, which is virtually identical for K and Na (0.25~e/\AA$^3$
and 0.27~e/\AA$^3$ respectively).

 \section{Sample Preparation}
The eutectic alloy was prepared by mixing liquid K and Na in an
inert atmosphere  glove box ($<$~2~ppm $  H_2O$, $<$1~ppm $ O_2$).
The alloy  was transferred into a stainless steel reservoir and
sealed with a Teflon O-ring. The design of the reservoir is
similar to the one used in recent Hg
experiments.\cite{mercurytemp} The reservoir is connected to a
stainless steel UHV valve which is attached to a UHV flange. This
assembly was connected to the UHV chamber, which was baked out at
150$^{\circ}$~C until a vacuum in the $10^{-10}$~Torr range
resulted. After reaching this pressure, the Mo sample pan (20~mm
diameter, 1~mm depth) in the UHV chamber was sputtered clean with
Ar$^+$ ions. Sputtering  removes the native Mo-oxide layer and
facilitates the wetting of the sample pan by the AM.  After
sputtering, the UHV valve to the reservoir was opened and the
liquid alloy metered into the sample pan. As we will show below,
our measurements under these conditions are consistent with a
surface free of oxide on an atomic level.

 \section{Experimental geometry}
Experiments were  carried out using the liquid surface
spectrometer at beamline X25  at the National Synchrotron Light
Source, operating with an x-ray wavelength $\lambda=$0.65~{\AA}
and a resolution of $\Delta q_z \approx 0.05$~{\AA}$^{-1}$. In the
x-ray reflectivity (XR) geometry, incident angle $\alpha$  and
reflected angle $\beta = \alpha$ are varied simultaneously with
$\beta$ detected within the plane of incidence. The background
intensity, due mainly to scattering from the bulk liquid, is
subtracted from the specular signal by displacing the detector out
of the reflection plane by slightly more than one resolution
width.\cite{indium,gaprb,masa} The specular reflectivity from the
surface, $R(q_z)$, is a function of the normal component, $q_z =
(4\pi/\lambda)\sin\alpha$, of the momentum transfer
$\vec{k}_{out}-\vec{k}_{in}=$ (0,0,$q_z$). $R (q_z)$ yields
information about surface roughness and  surface-normal structure
and may be approximated as\cite{indium}
\begin{equation}
    R(q_z) =  R_f(q_z)
        \left| \Phi (q_z) \right| ^2
\left(\frac{q_{res}}{q_{max}}\right)^{\eta} \label{kna_eq:circle}
\end{equation}
where $R_f (q_z)$ is the Fresnel reflectivity of a flat,
laterally homogeneous surface.
The surface structure factor,  $\Phi (q_z)$,
is the Fourier transform of the gradient of the intrinsic
surface-normal  density profile.  The power-law term
with
$\eta=(k_BT/2\pi\gamma) q_z^2$  accounts
for roughening of the intrinsic density profile
by capillary waves (CW).
Scattering of x-ray photons by CW  outside of the detector
resolution $q_{res}$ result in a loss in observed
intensity. The short-wavelength cutoff $q_{max}$
is determined by the particle size $a$ with
$q_{max} \approx \pi/a$.
The observed reflectivity is reduced when either
 the temperature $T$  is increased
or the surface tension $\gamma$ is reduced.

For off-specular diffuse scattering (DS), the incoming angle
$\alpha$ is kept constant while  $\beta$ is varied, straddling the
specular condition (see inset of Figure~\ref{kna_fig:phase}). The
bulk diffuse background  has to be subtracted from the measured
overall intensity. An acurate description of the theoretical
diffuse scattering requires inclusion of the contribution of the
surface to the bulk background scattering.\cite{indium,masa} The
lineshape of the DS scans is determined principally by the
power-law term in Eq.~\ref{kna_eq:circle}.  Analysis of DS gives
access to in-plane correlations of the surface and surface
inhomogeneities that might be present.\cite{masa} Simultaneous
analysis of XR and DS allows for a consistent determination of
surface roughness and structure. As has been pointed out
recently,\cite{petersrn} it is only for $\eta < 2$ that the
reflected intensity displays the cusp-like singularity centered at
the specular condition $\alpha = \beta$. For $\eta \geq 2$, it is
not possible to distinguish surface scattering from bulk diffuse
scattering since both are dominated by correlations of the same
length scale. The value of $q_z$ at which $\eta$ reaches the limit
of 2  at room temperature is about 1.7~\AA$^{-1}$ for
K$_{67}$Na$_{33}$.

\section{X-Ray Diffuse Scattering Measurements}
\begin{figure}[tbp]
\epsfig{file=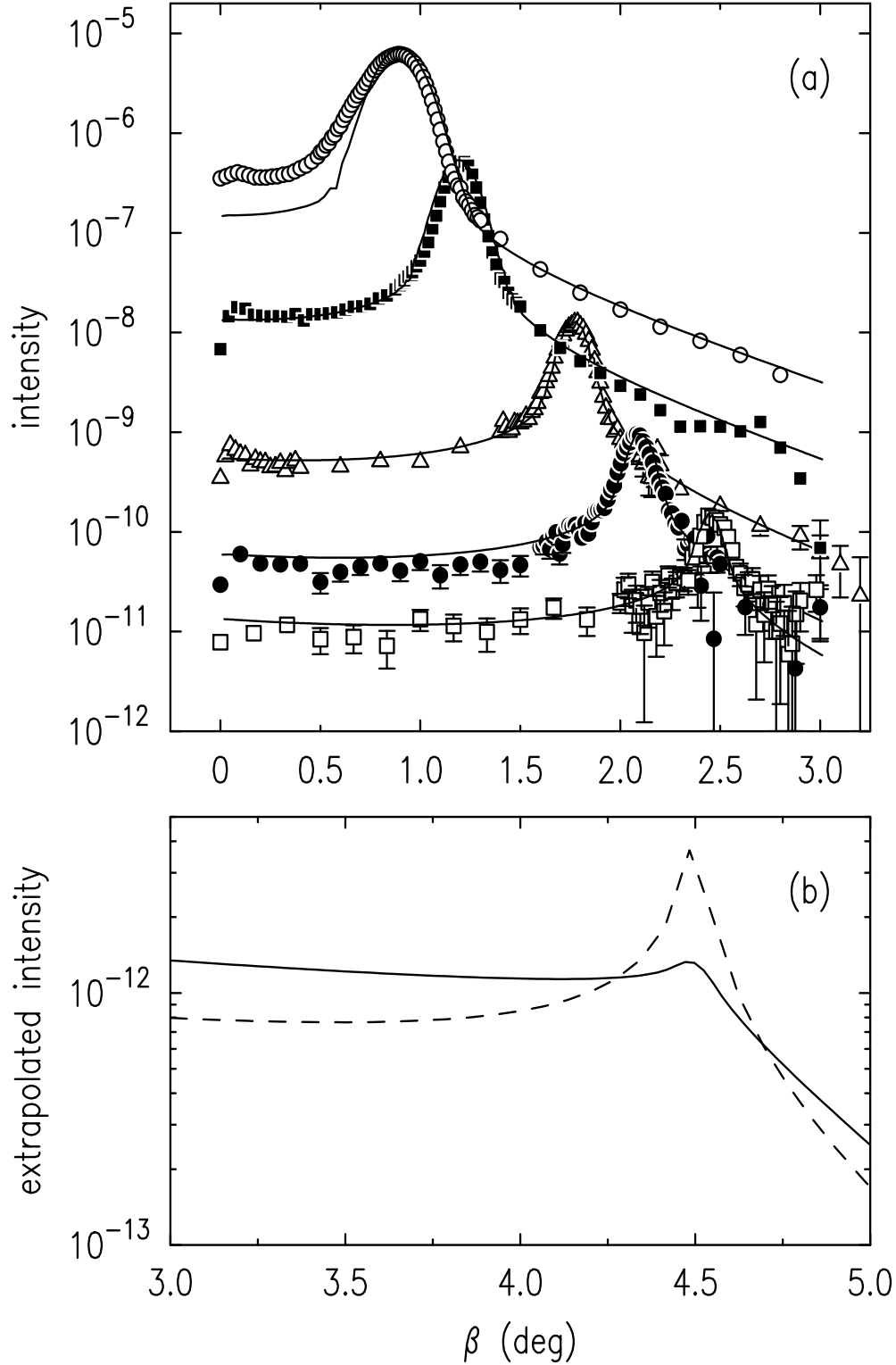, angle=0, width=0.8\columnwidth} \caption{
(a) Off-specular diffuse scattering from the surface of  liquid $
K_{67}Na_{33}$ for different incoming angles $\alpha$:
(\protect$\circ$): 0.9$ ^{\circ}$ (multiplied by 25 for clarity),
(\protect\rule{1.5mm}{1.5mm}): 1.2$  ^{\circ}$ ($\times$ 5),
 (\protect$\triangle$): 1.77$  ^{\circ}$,
(\protect$\bullet$): 2.1$  ^{\circ}$ ($\times$ .2),
(\protect$\Box$): 2.45$  ^{\circ}$ ($\times$ .1). (b) Extrapolated
diffuse scattering (assuming no surface layering) for $\alpha =$
4.5~$  ^{\circ}$: (---) resolution of the present experiment
($\Delta q_z \approx$ 0.03~{\AA}$^{-1}$). ( - - - ) assuming the
resolution attainable at a low divergence x-ray source: ($\Delta q_z
\approx 3 \times 10^{-4}$~{\AA}$^{-1}$). (intensity $\times$ 100).
}\label{kna_fig:diffuse}

\end{figure}

Diffuse scattering measurements from the surface of $
K_{67}Na_{33}$ taken at different incoming angles $\alpha$
(Figure~\ref{kna_fig:diffuse}(a)) show that the liquid alloy has a
uniform surface roughened by capillary waves. The solid lines
through the data are calculated from CW theory (power-law term in
Eq.~\ref{kna_eq:circle}) with the surface roughness  entirely due
to thermally activated surface waves. This model has no adjustable
parameters, and the surface tension of 0.110$\pm0.003$~N/m is
consistent with macroscopic measurements.\cite{tension} The
agreement between the data and CW theory is excellent. The only
limitation is in the low-angle region ($\alpha=0.9^\circ$ and
$\beta < \alpha$), where the footprint of the incoming beam is
larger than the flat part of the sample. Apart from this
deviation, the sample shows no excess scattering which would be
expected from an inhomogeneous surface.\cite{indium,masa} The
absence of excess diffuse scattering indicates that  the alloy
surface is free of microscopic patches or islands of possible
contaminants such as oxide or hydride.

 \section{X-Ray Reflectivity Measurements}
\begin{figure}[tbp]
\epsfig{file=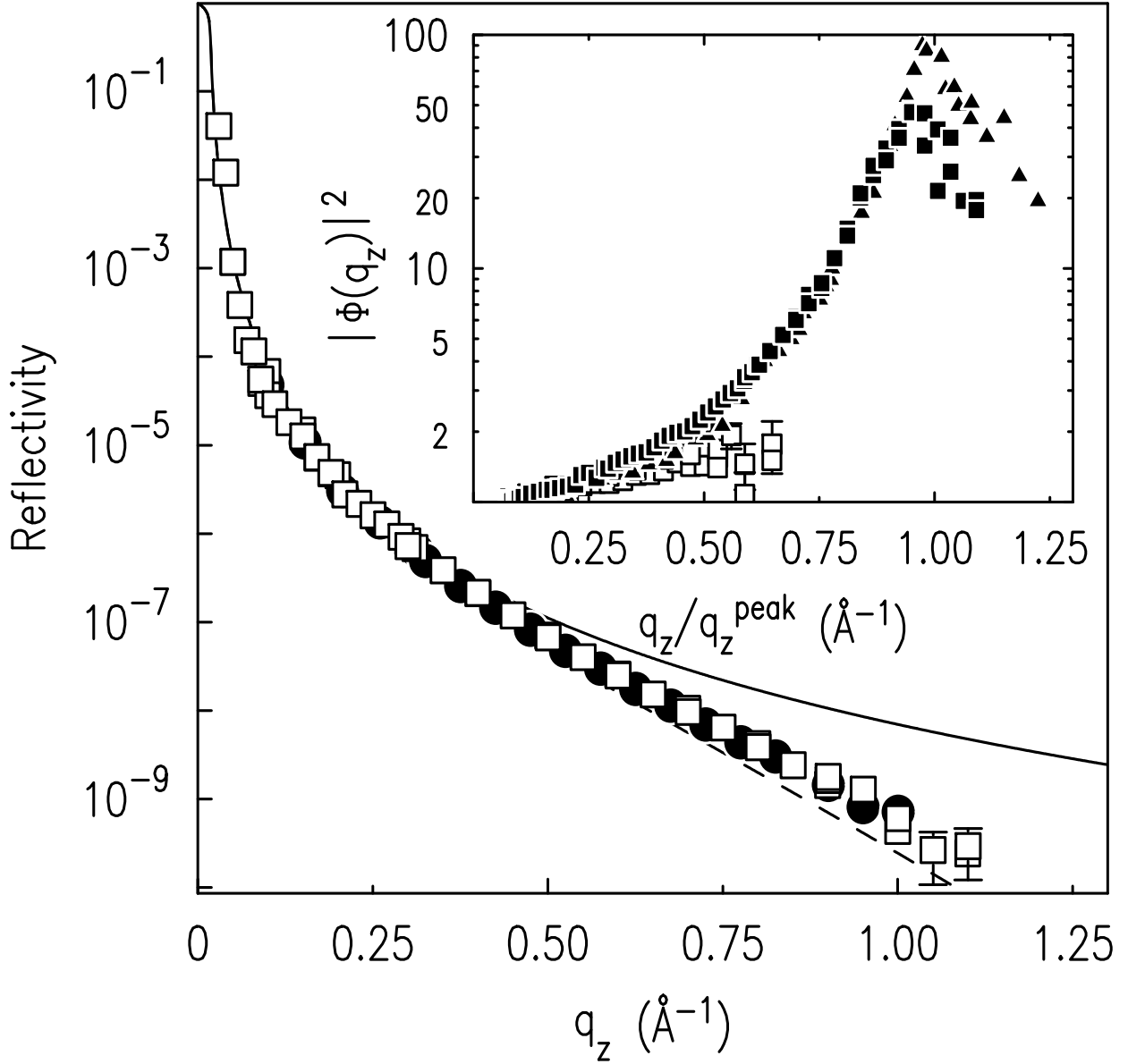, angle=0, width=1.0\columnwidth}
\caption{X-ray reflectivity from a clean liquid $  K_{67}Na_{33}$
surface at room temperature ($\Box$) and $  K_{67}Na_{33}$
following an exposure to 3000~L of oxygen ($\bullet$) Also
included for reference: (---) calculated Fresnel reflectivity
 from an ideal $  K_{67}Na_{33}$ surface; (- - -) calculated
reflectivity $R_{cw}$ from a $  K_{67}Na_{33}$ surface roughened by
capillary waves at room temperature. Inset: Surface structure factor
$|\Phi(q_z)|^2$ of liquid $  K_{67}Na_{33}$ at room temperature
($\Box$). For comparison the surface structure factors for liquid Ga
(closed triangles) and In (closed boxes) are included. $q_z$ is
normalized to the position of the first peak in the bulk $S(q)$.}
\label{kna_fig:reflectivity}
\end{figure}

As a result of this excellent agreement between CW theory and DS,
the reflectivity from the alloy surface shown in
Figure~\ref{kna_fig:reflectivity} (open squares) can be compared
to the calculated
 reflectivity  from a K$_{67}$Na$_{33}$ surface
which is roughened by CW but displays no structure (i.e.
$\Phi(q_z)=1$). The dashed line in
Figure~\ref{kna_fig:reflectivity} displays this theoretical
 prediction and it is obvious
that the data rise above it, consistent with the low-angle
behavior of surface induced layering observed
previously.\cite{gallium,indium} exist. This point is emphasized
further in the inset where the measured $R(q_z)$ is divided by
$R_f$ and the power-law CW term in Eq.~\ref{kna_eq:circle} to
obtain a direct measure of the surface structure factor
$\Phi(q_z)$.  As can be seen, the value of $\Phi(q_z)$ rises above
unity with increasing $q_z$ as expected when constructive
interference of x-rays due to surface  layering is present. The
low angle tail of the surface structure factor of
K$_{67}$Na$_{33}$   is now compared with that of liquid Ga and In
reported previously .\cite{gallium,indium} To normalize $q_z$ to
the atomic size, $q_z$ has been divided by $q_z^{max}$ of the
first maximum of $S(q)$. For K$_{67}$Na$_{33}$, $q_z^{max}\approx
1.7$~{\AA}$^{-1}$ whereas for Ga and In $q_z^{max}\approx
2.4$~{\AA}$^{-1}$. It is evident that the surface structure factor
indicative of surface-normal layering  does not increase as
quickly for  K$_{67}$Na$_{33}$ as for  Ga or In. Assuming that the
same layering model that successfully describes the surface of
Ga\cite{gallium} and In\cite{indium} also applies to the KNa alloy
the measurement indicates that either the surface layering of the
alloy is significantly weaker than that of both Ga or In, or the
length over which the layering decays into the bulk is
signficantly shorter. One can speculate that there might be a
correlation between these layering length scales and the surface
tension. This would mean that weak or quickly decaying layering is
expected for LM with exceptionally low surface tension. Clearly,
data extending well beyond $q_z = 1$~\AA$^{-1}$ are needed to
support this conclusion.

 \section{Capillary Wave Contributions}
Several factors contribute to the difficulty of extending
reflectivity measurements to values of $q_z$ large enough to cover
the expected position of the layering peak. The reflected
intensity falls over almost ten orders of magnitude, while the
background from the bulk structure factor increases, as this value
of $q_z$ is approached. We must also consider that the exponent
$\eta$ approaches its limiting value of 2 for $q_z \rightarrow
1.7$~{\AA}$^{-1}$ so that the specular  cusp becomes broadened and
ultimately lost in the diffuse background. Because of these
effects, simply increasing the photon flux is not sufficient to
improve the measurement substantially. A better approach is to
significantly improve the resolution. This is illustrated in
Figure~\ref{kna_fig:diffuse}(b), where the calculated DS for
$\alpha = 4.5$~degrees (corresponding to $q_z = 1.5$~{\AA}$^{-1}$)
is shown for the current experimental resolution (solid line) and
for a resolution sharper by a factor of 200 (dashed line).
Although the intensity is much reduced in the latter case, the
peak is much better defined. Reflectivity measurements conducted
at  low divergence synchrotron x-ray sources that are presently
becoming available may make such extensions of the $q_z$ range
possible.

 \section{Oxidation Studies}
To explore the effect of possible contaminants on the surface
structure, we deliberately exposed the liquid alloy surface to
controlled amounts of pure oxygen and hydrogen gas, measured in
units of monolayer surface coverage (Langmuir).\cite{galliumoxide}
The liquid alkali surface has been exposed to up to 3000~L oxygen
and up to 6000~L hydrogen. As an example,$R(q_z)$ at the maximum
exposure of O$_2$ is shown as filled circles in
Figure~\ref{kna_fig:reflectivity}. In no case was a measurable
effect on $R(q_z)$ observed, indicating no change in the
surface-normal structure. This insensitivity of the liquid
K$_{67}$Na$_{33}$ surface to small amounts of oxygen is
 in striking contrast to similar experiments on
liquid Ga\cite{galliumoxide} and Hg\cite{lam} where the formation
of a thin homogeneous oxide film has been observed upon exposure
of the liquid surface to as little as 200~L of oxygen. This
difference is due to the fact that the low surface tension of
liquid AM precludes segregation of oxide from the bulk. By
contrast, high surface tension LM such as Ga or Hg lower their
high overall surface energy by segregation of a contaminant  film.
These results are also in agreement with photoelectron
spectroscopy experiments requiring a similarly clean liquid AM
surface.\cite{oelhafen}

 \section{Summary}
In summary, we have shown that it is possible to investigate the
surface structure of an atomically clean liquid AM under UHV
conditions.  Exposing the sample to up to 6000~L of H$_2$ or O$_2$
has no effect on the atomic surface structure since the
exceptionally low surface tension of AM precludes contaminant
segregation. From DS we extract the surface roughness, which
agrees well with capillary wave theory, and with the macroscopic
surface tension reported in the literature. The $q_z$-dependence
of the
 surface structure factor determined from the XR
is consistent with  constructive interference of x-rays due to
surface-normal layering. This surface-induced layering in liquid $
K_{67}Na_{33}$ appears to be  much weaker than that found in high
surface tension LM such as In or Ga.


\begin{thebibliography}{1}
\bibitem{peak} The position of  the layering peak
is expected to be close to that of
the first peak of the bulk $S(q)$.
\end{thebibliography}
\end{document}